\documentclass[aps]{revtex4}
\usepackage[colorlinks=true, pdfstartview=FitV, linkcolor=blue, citecolor=red, urlcolor=magenta, breaklinks=true]{hyperref}
\usepackage{graphicx}  
\usepackage{amsfonts}
\begin{document}
\title{The entropy of the noncommutative acoustic black hole based on generalized uncertainty principle}
\author{M. A. Anacleto, F. A. Brito, E. Passos and W. P. Santos}
\email{anacleto, fabrito, passos@df.ufcg.edu.br}
\affiliation{Departamento de F\'{\i}sica, Universidade Federal de Campina Grande, Caixa Postal 10071, 58109-970 Campina Grande, Para\'{\i}ba, Brazil}
\begin{abstract} 

In this paper we investigate statistical entropy of a 3-dimensional rotating acoustic black  hole based on generalized uncertainty principle. In our results  we obtain an area entropy and a correction term associated with the  noncommutative acoustic black hole
when $ \lambda $ introduced in the generalized uncertainty principle takes a specific value. 
However, in this method, it is not needed to introduce the ultraviolet cut-off and divergences are eliminated. Moreover, the small mass approximation is not necessary in the original brick-wall model.

\end{abstract}
\maketitle
\pretolerance10000

\section{Introduction}
The concept of acoustic black holes was proposed in 1981 by Unruh~\cite{Unruh} and has been extensively studied in the literature~\cite{MV, Volovik, others}. The connection between black hole physics and the theory of supersonic acoustic flow is now well established and has been developed to investigate the Hawking radiation and other phenomena for understanding quantum gravity. Acoustic black holes was found to possess many of the fundamental properties of black holes in general relativity. Thus, many fluid systems have been investigated on a variety of analog models of acoustic black holes, including gravity wave~\cite{RS}, water~\cite{Mathis}, slow light~\cite{UL}, optical fiber~\cite{Philbin} and  electromagnetic waveguide~\cite{RSch}. The models of superfluid helium II~\cite{Novello}, atomic Bose-Einstein condensates~\cite{Garay,OL} and one-dimensional Fermi degenerate noninteracting gas~\cite{SG} have been proposed to create an acoustic black hole geometry in
the laboratory. A relativistic version of acoustic black holes has been presented in~\cite{Xian,ABP}.

Recently, in Ref.~\cite{Rinaldi} was investigated (1 + 1)-dimensional acoustic black hole entropy by the brick-wall method. In order to obtain a finite result, they had to introduce the ultraviolet cut-off. So their calculation suggested that analog black hole entropy has the “cut-off problem” similar to that of gravitational black hole entropy. More recently in \cite{Rinaldi:2011aa} the author uses transverse modes in order to cure the divergences.

The study on the statistical origin of black hole entropy has been extensively explored by several authors --- see for instance ~\cite{Wilczek}. 
The brick-wall method proposed by G. 't Hooft has been used for calculations on the black hole, promoting the understanding of the origin of black hole entropy. According to G. 't Hooft, black hole entropy is just the entropy of quantum fields outside the black hole horizon. However, when one calculates the black hole statistical entropy by this method, to avoid the divergence of states density near black hole horizon, an ultraviolet cut-off must be introduced. The statistical entropy of various black holes has been calculated via corrected state density of the generalized uncertainty principle (GUP)~\cite{XLi}. Thus, the results show that near the horizon quantum state density and its statistical entropy are finite. In~\cite{KN} a relation for the corrected states density by GUP has been proposed
\begin{eqnarray}
dn=\frac{d^3 x d^3 p}{(2\pi)^3}e^{-\lambda p^2},
\end{eqnarray}
where $p^2=p^{i}p_{i}  $, and $ \lambda$ { plays the role of the Planck scale in a fluid at high energy regimes --- See below}.

In~\cite{Zhao} using a new equation of state density due to GUP, the statistical entropy of a 3-dimensional rotating acoustic black hole has been analyzed. It was shown that  using the quantum statistical method the entropy of the rotating acoustic black hole was calculated, and the Bekenstein-Hawking area entropy of acoustic black hole and its correction term was obtained. Therefore, considering the effect due to GUP on the equation of state density,  no cut-off is needed and the divergence in the brick-wall model disappears.

In this paper, we apply the acoustic black hole metrics obtained from a relativistic fluid in a noncommutative spacetime~\cite{ABP12} via the Seiberg-Witten map to study the entropy of the rotating acoustic black hole.
Whereas, on one hand our objective is to see if using an equation of state of the GUP divergences are eliminated as in the gravitational case, furthermore, we wonder whether the noncommutativity of the spacetime affects the GUP itself. 
{This is also motivated by the fact that in high energy physics both strong spacetime noncommutativity and quark gluon plasma (QGP) may take place together. Thus, it seems to be natural to look for acoustic black holes in a QGP fluid with spacetime noncommutativity  in this regime. Acoustic phenomena in QGP matter can be seen in Ref.~\cite{shk} and acoustic black holes in a plasma fluid can be found in Ref.~\cite{BH-plasma}.

Differently of the most cases studied, we consider the acoustic black hole metrics obtained from a relativistic fluid in a noncommutative spacetime. 
 The effects of this
set up is such that the fluctuations of the fluids are also affected.  The sound waves inherit spacetime noncommutativity of the fluid and may lose the Lorentz invariance. As a consequence the Hawking temperature is directly affected by the spacetime noncommutativity. Analogously to Lorentz-violating gravitational black holes \cite{syb,adam}, the effective Hawking temperature of the noncommutativity  acoustic black holes 
now is {\it not} universal for all species of particles. It depends on the maximal attainable velocity of this species.
Furthermore, the acoustic black hole metric can be identified with 
an acoustic Kerr-like black hole. It was found in~\cite{ABP12} that  the spacetime noncommutativity  affects the rate of loss of mass of the black hole.  Thus for suitable values of
the spacetime noncommutativity parameter a wider or narrower spectrum of particle wave function can be scattered with increased amplitude by the acoustic black hole. This
increases or decreases the superressonance phenomenon previously studied in \cite{Basak:2002aw,SBP}.
}

In our study we shall focus on the quantum statistical method  to determine the entropy of an acoustic black hole using the equation of state density from the GUP. We anticipate that we have obtained the Bekenstein-Hawking entropy of acoustic black hole and its correction term are obtained via the quantum statistical method. There is no need to introduce the ultraviolet cut-off  and divergences are eliminated.

\section{The Acoustic Metric in Noncommutative Abelian Higgs Model}
\label{II}
{In this section we consider the noncommutative version of the Abelian Higgs model {in (3+1) dimensions}. 
The noncommutativity is introduced by modifying its scalar and gauge sector by replacing the usual product of fields by the Moyal product \cite{SW,SGhosh,rivelles,revnc} --- see also \cite{Cai:2007xr,Cai:2014hja} for related issues.}
Thus, the Lagrangian of the noncommutative Abelian Higgs model in flat space is
\begin{eqnarray}
\label{eqAHM}
\hat{\cal L}&=&-\frac{1}{4}\hat{F}_{\mu\nu}\ast\hat{F}^{\mu\nu} 
+(D_{\mu}\hat{\phi})^{\dagger}\ast D^{\mu}\hat{\phi}+ m^2\hat{\phi}^{\dagger}\ast\hat{\phi}-b\hat{\phi}^{\dagger}\ast\hat{\phi}\ast\hat{\phi}^{\dagger}\ast\hat{\phi},
\end{eqnarray}
where the hat indicates that the variable is noncommutative and the $ \ast $-product is the so-called Moyal-Weyl product or star product which is defined in terms of a real antisymmetric matrix $ \theta^{\mu\nu}$ that
parameterizes the noncommutativity of Minkowski spacetime
\begin{eqnarray}
[x^{\mu},x^{\nu}]=i\theta^{\mu\nu}, \quad \mu,\nu=0,1,\cdots,D-1.
\end{eqnarray}
The $ \ast $-product for two fields $f(x)$ and $g(x)$ is given by
\begin{eqnarray}
f(x)\ast g(x)=\exp\left(\frac{i}{2}\theta^{\mu\nu}\partial^{x}_{\mu}\partial^{y}_{\nu}\right)f(x)g(y)\vert_{x=y}.
\end{eqnarray}
In (\ref{eqAHM}) the noncommutative  fields can be expanded in a formal series in $ \theta $. 
{ As one knows the parameter $\theta^{\alpha\beta}$ is a constant, real-valued antisymmetric $D\times D$- matrix in $D$-dimensional spacetime with dimensions of length squared. For a review see \cite{revnc}. }
Using the  Seiberg-Witten (SW)  map this expansion can be constructed in terms of the original fields of a commutative theory transforming under the ordinary transformation laws.

Now using the Seiberg-Witten map \cite{SW}, up to the lowest order in the spacetime noncommutative parameter $\theta^{\mu\nu}$, we find 
\begin{eqnarray}
&&\hat{A}_{\mu}=A_{\mu}+\theta^{\nu\rho}A_{\rho}(\partial_{\nu}A_{\mu}-\frac{1}{2}\partial_{\mu}A_{\nu}),
\nonumber\\
&&\hat{F}_{\mu\nu}=F_{\mu\nu}+\theta^{\rho\beta}(F_{\mu\rho}F_{\nu\beta}+A_{\rho}\partial_{\beta}F_{\mu\nu}),
\nonumber\\
&&\hat{\phi}=\phi-\frac{1}{2}\theta^{\mu\nu}A_{\mu}\partial_{\nu}\phi.
\end{eqnarray}
{ This very useful map allows us to study noncommutative effects in the framework of
commutative quantum field theory. 

Thus the corresponding theory in a commutative spacetime { in (3+1) dimensions} is~\cite{SGhosh}}
\begin{eqnarray}
\label{acao}
\hat{\cal L}&=&-\frac{1}{4}F_{\mu\nu}F^{\mu\nu}\left(1+\frac{1}{2}\theta^{\alpha\beta}F_{\alpha\beta}\right) 
+\left(1-\frac{1}{4}\theta^{\alpha\beta}F_{\alpha\beta}\right)\left(|D_{\mu}\phi|^2+ m^2|\phi|^2-b|\phi|^4\right)
\nonumber\\
&+&\frac{1}{2}\theta^{\alpha\beta}F_{\alpha\mu}\left[(D_{\beta}\phi)^{\dagger}D^{\mu}\phi+(D^{\mu}\phi)^{\dagger}D_{\beta}\phi \right],
\end{eqnarray}
where $F_{\mu\nu}=\partial_{\mu}A_{\nu}-\partial_{\nu}A_{\mu}$ and  $D_{\mu}\phi=\partial_{\mu}\phi - ieA_{\mu}\phi$.

{Let us briefly review the steps to find the noncommutative acoustic black hole metric { in (3+1) dimensions} from quantum field theory. Firstly, we decompose the scalar field as  $\phi = \sqrt{\rho(x, t)} \exp {(iS(x, t))}$ into the original Lagrangian to find
\begin{eqnarray}
{\cal L}&=&-\frac{1}{4}F_{\mu\nu}F^{\mu\nu}\left(1-2\vec{\theta}\cdot\vec{B}\right)
+\rho(\tilde{\theta}g^{\mu\nu}+\Theta^{\mu\nu}){\cal D}_{\mu}S{\cal D}_{\nu}S+\tilde{\theta} m^2\rho-\tilde{\theta}b\rho^2
+\frac{\rho}{\sqrt{\rho}}(\tilde{\theta}g^{\mu\nu}+\Theta^{\mu\nu})\partial_{\mu}\partial_{\nu}\sqrt{\rho},
\end{eqnarray}
where ${\cal D}_{\mu}=\partial_{\mu}-eA_{\mu}/S  $, $\tilde{\theta}=(1+\vec{\theta}\cdot\vec{B})$, $\vec{B}=\nabla\times\vec{A}$ and $\Theta^{\mu\nu}=\theta^{\alpha\mu}{F_{\alpha}}^{\nu}$. 
In our calculations we consider the case where there is no noncommutativity between space and time, that is $\theta^{0i}=0$ and use $\theta^{ij}=\varepsilon^{ijk}\theta^{k}$, $F^{i0}=E^{i}$ and $F^{ij}=\varepsilon^{ijk}B^{k}$.

Secondly, linearizing the equations of motion around the background $(\rho_0,S_0)$, with $\rho=\rho_0+\rho_1$ and  $S=S_0+\psi$ we find the equation of motion for a linear acoustic disturbance $\psi$ given by a Klein-Gordon equation in a curved space
\begin{eqnarray}
\frac{1}{\sqrt{-g}}\partial_{\mu}(\sqrt{-g}g^{\mu\nu}\partial_{\nu})\psi=0,
\end{eqnarray}
where $g_{\mu\nu}$ just represents the acoustic metrics { in (3+1) dimensions}. { We should comment that in our previous computation we assumed linear perturbations just in the scalar sector,  whereas the vector field $A_\mu$ remain unchanged.}
 }

In the following we shall focus on the planar rotating acoustic noncommutative black hole metrics { in (2+1) dimensions}~\cite{ABP12} to address the issues of the entropy of three-dimensional rotating acoustic black hole. For the sake of simplicity, we shall consider {\it two types} of a noncommutative spacetime medium by choosing first pure magnetic sector and then we shall focus on the  pure electric sector.

\subsection{The case $B\neq 0$ and $E=0$}
The acoustic line element in polar coordinates on the noncommutative plane { in (2+1) dimensions}, up to an irrelevant position-independent factor, in the nonrelativistic limit { was obtained in \cite{ABP12} and is given by }
\begin{eqnarray}
ds^2&=&-[(1-3\theta_{z}B_{z})c^{2}-(1+3\theta_{z}B_{z})(v^2_{r}+v^2_{\phi})]dt^2
-2(1+2\theta_{z}B_{z})(v_{r}dr+v_{\phi}rd{\phi})dt
\nonumber\\
&+&(1+\theta_{z}B_{z})(dr^2+r^2d\phi^2).
\end{eqnarray}
{where $B_z$ is the magnitude of the magnetic field in the $z$ direction, $ \theta_z $ is the noncommutative parameter, 
$c$ is the sound velocity in the fluid and $v$ is the fluid velocity.}
We consider the flow with the velocity potential $\psi(r,\phi) = A\ln{r} + B\phi$  whose velocity profile in polar coordinates on the plane is  given by
\begin{eqnarray}
\vec{v}=\frac{A}{r}\hat{r}+\frac{B}{r}\hat{\phi},
\end{eqnarray}
where $B$ and $A$ are the constants of circulation and draining rates of the fluid flow.

Let us now consider the transformations of the time and the azimuthal angle coordinates as follows 
\begin{eqnarray}
&&d\tau=dt+\frac{(1+2\theta_{z}B_{z})Ardr}
{[(1-3\theta_{z}B_{z})c^2r^2-(1+3\theta_{z}B_{z})A^2]},
\nonumber\\
&&d\varphi=d\phi+\frac{ABdr}{r[c^2r^2-A^2]}.
\end{eqnarray}
In these new coordinates the metric becomes
\begin{eqnarray}
\label{ELB}
ds^2\!=\!\tilde{\theta}
\left[-(1-4\Theta)\left(1-\frac{(1+6\Theta)(A^2+B^2)}{c^2r^2}\right)d\tau^2
+\left(1-\frac{(1+6\Theta)A^2}{c^2r^2}\right)^{-1}dr^2
-\frac{2{\tilde\theta}B}{c}d\varphi d\tau+r^2d\varphi^2\right],
\end{eqnarray}
where $\Theta=\theta_{z}B_{z}$ and $\tilde{\theta}=1+\Theta$. 
The metric can be now written in the form
\begin{eqnarray}
g_{\mu\nu}={\tilde\theta}\left[\begin{array}{clcl}
-(1-4\Theta)\left[1-\frac{r_{e}^2}{r^2}\right] &\quad\quad 0& -\frac{{\tilde\theta}B}{c}\\
0 & \left(1-\frac{r_{h}^2}{r^2} \right)^{-1}& 0\\
-\frac{{\tilde\theta}B}{c} &\quad\quad 0 & r^2
\end{array}\right].
\end{eqnarray}

The radius of the ergosphere is given by $g_{00}(r_{e}) = 0$, whereas the horizon is given by the coordinate singularity $g_{rr}(r_{h}) = 0$, that is
\begin{eqnarray}
r_{e}=\sqrt{r_{h}^2+\frac{(1+6\Theta)B^2}{c^2}}, \quad r_{h}=\frac{(1+6\Theta)^{1/2}|A|}{c}.
\end{eqnarray}
{Before going further let us investigate curvature singularities. Let us do this by checking the invariants $R$ and $R_{\mu\nu\lambda\sigma}R^{\mu\nu\lambda\sigma}$.  They are computed through the metric  (\ref{ELB}). By power expanding them in $\Theta$ we find
\begin{eqnarray}\label{curvature-invariants}
R=\frac{-2(A^2+B^2)(1+6\Theta)}{r^4}+{\cal O}\left(\frac{\Theta^2}{(r^2-A^2)^2r^4}\right), \; R_{\mu\nu\lambda\sigma}R^{\mu\nu\lambda\sigma}=\frac{44(A^2+B^2)^2(12+\Theta)}{r^8}+{\cal O}\left(\frac{\Theta^2}{(r^2-A^2)^2r^8}\right),
\end{eqnarray}
where we have assumed $c=1$ for simplicity. Notice that both invariants have no curvature singularities at $r\neq0$ up to linear analysis. However, we find a singularity at $r=A$ for quadratic (and higher order) terms in $\Theta$. Since we are considering a linear theory in the non-commutativity parameter $\theta^{\mu\nu}$ from the beginning, the singularity at $r=A$ that relies only on higher order terms in $\Theta\equiv\theta_z B_z$ should be disregarded for consistence. 
}

Now we obtain the Hawking temperature of the acoustic
black hole as 
\begin{eqnarray}
T_{h}=\frac{k}{2\pi}=\frac{\left(1-2\Theta\right)c^2}{2\pi r_{h}}.
\end{eqnarray}
While the Unruh temperature for an observer at a distance $r$ is
\begin{eqnarray}
T=\frac{a}{4\pi}=\frac{f^{\prime}(r_h)}{4\pi}F^{-1/2}(r).
\end{eqnarray}
They satisfy the following relation
\begin{eqnarray}
T_h=\sqrt{F(r)}T=\frac{f^{\prime}(r_h)}{4\pi}.
\end{eqnarray}
where
\begin{eqnarray}
f(r)=(1-2\Theta)\left(1-\frac{r^2_h}{r^2} \right), \quad\quad F(r)=\frac{g_{tt}g_{\varphi\varphi}-g^2_{t\varphi}}{g_{\varphi\varphi}}=(1-5\Theta)\left(1-\frac{r^2_h}{r^2} \right).
\end{eqnarray}

\section{The statistical entropy}
The partition function for a Bose system is
\begin{eqnarray}
\ln Z_0=-\sum_i g_i\Big ( 1-e^{-\beta\epsilon_i}  \Big),
\end{eqnarray}
and the area element with constant time $t$ coordinate is
\begin{eqnarray}
ds=2\pi\sqrt{g_{\varphi\varphi} g_{rr}}dr.
\end{eqnarray}
The partition function of the system  outside the acoustic black hole horizon is given by
\begin{eqnarray}
\ln Z&=&-\int 2\pi\sqrt{g_{\varphi\varphi} g_{rr}}dr \sum_i g_i\Big ( 1-e^{-\beta\epsilon_i}  \Big)
\nonumber\\
&=&-\int\sqrt{g_{\varphi\varphi} g_{rr}}dr \int_0^{\infty} dp \Big(pe^{-\lambda p^2}\Big)\Big ( 1-e^{-\beta\omega_0}  \Big)
\nonumber\\
&\approx&\int\sqrt{g_{\varphi\varphi} g_{rr}}dr \int^{\infty}_{m\sqrt{-\tilde{g}_{tt}}} 
\frac{\beta_0 e^{-\lambda p^2}p^2 d\omega}{2\Big (e^{\beta\omega_0} -1 \Big)},
\end{eqnarray}
where $ \beta=\beta_0\sqrt{-\tilde{g}_{tt}} $, $\omega=\omega_0\sqrt{-\tilde{g}_{tt}}$ and $-\tilde{g}_{tt}=-\frac{g_{tt}g_{\varphi\varphi}-g^2_{t\varphi}}{g_{\varphi\varphi}}$.
According to the relation between the free energy and partition
function, we can derive the free energy of the system as
\begin{eqnarray}
F=-\frac{1}{\beta_0}\ln Z=\int\sqrt{g_{\varphi\varphi} g_{rr}}dr \int^{\infty}_{m\sqrt{-\tilde{g}_{tt}}} 
\frac{ e^{-\lambda p^2}p^2 d\omega}{2\Big (e^{\beta\omega_0} -1 \Big)},
\end{eqnarray}
and the entropy of the system is
\begin{eqnarray}\label{entropia}
S&=&\beta_0^2\frac{\partial F}{\partial\beta_0}
=\beta_0^2\int\sqrt{g_{\varphi\varphi} g_{rr}}dr \int^{\infty}_{m\sqrt{-\tilde{g}_{tt}}} 
\frac{\omega e^{\beta\omega_0} e^{-\lambda p^2}p^2 d\omega}{2\Big (e^{\beta\omega_0} -1 \Big)^2}
\nonumber\\
&=&\frac{1}{2}\int\sqrt{g_{\varphi\varphi} g_{rr}}dr \int^{\infty}_{m\beta} 
\frac{x e^{x} }{(e^{x} -1 )^2}e^{-\lambda\Big(\frac{x^2}{\beta^2}-m^2\Big)}\Big(\frac{x^2}{\beta^2}-m^2 \Big) dx,
\end{eqnarray}
where we have defined $ x=\beta\omega_{0}=\beta_0\omega $, and we have utilized the relation among energy, momentum 
and mass $\frac{\omega^2}{-\tilde{g}_{tt}}=\frac{x^2}{\beta^2}=p^2+m^2$, being $m$ the static mass of particles. 
Thus, we integrate (\ref{entropia}) with respect to $r$ near the black hole horizon. 
Near the horizon, $ \tilde{g}_{tt}(r_h)\rightarrow 0 $, so we have
\begin{eqnarray}
S&=&\frac{1}{2}\int\sqrt{g_{\varphi\varphi} g_{rr}}dr \int^{\infty}_{0} 
\frac{x^3 e^{x} }{\beta^2(e^{x} -1 )^2}e^{-\lambda\frac{x^2}{\beta^2}} dx
=\frac{1}{2\beta_0^2} \int^{\infty}_{0} 
\frac{ dx}{4\sinh^2(x/2)}I(x,\epsilon),
\end{eqnarray}
where
\begin{eqnarray}
I(x,\epsilon)=\int\frac{\sqrt{g_{\varphi\varphi} g_{rr}}}{-\tilde{g}_{tt}}x^3e^{-\lambda\frac{x^2}{\beta^2}}dr
=\int(1-4\Theta)\frac{x^3}{N^3}e^{-\lambda\frac{x^2}{\beta^2}}rdr, \quad\quad N^2=1-\frac{r^2_h}{r^2}
\end{eqnarray} 
Since we only consider the quantum field near the black hole horizon,	we take $[r_h , r_h	+\epsilon]$	as	the integral interval with respect to $r$, where $\epsilon$ is a positive small constant. 
When $r \rightarrow r_h$ , $N^2(r)\approx 2\kappa (r-r_h) $, so we have
\begin{eqnarray}
I(x,\epsilon)
=\int_{r_h}^{r_h +\epsilon}\frac{(r-r_h)+r_h}{[2\kappa(r-r_h)]^{3/2}}{x}^3 
e^{-\tilde{\lambda} {x}^2/[2\kappa(r-r_h)\beta^2_0]}dr, 
\end{eqnarray}
{where $\tilde{\lambda}=\lambda(1-5\Theta)^{-1}  $ } and $ \kappa=2\pi\beta_0^{-1}$ is the surface gravity of acoustic black hole horizon and by variable substitution $ t=\frac{\tilde{\lambda} x^2}{4\pi(r-r_h)\beta_0} $, we have
\begin{eqnarray}
I(x,\epsilon)
&=&(1-4\Theta)\int_{\delta}^{\infty}\left[\frac{\beta_0 {x}^4\sqrt{\tilde{\lambda}}}{(4\pi)^2}t^{-3/2} 
+\frac{ r_h\beta_0^2 {x}^2}{4\pi\sqrt{\tilde{\lambda}}}t^{-1/2}\right]e^{-t}dt
\nonumber\\
&=&(1-4\Theta)\left[\frac{\beta_0 {x}^4\sqrt{\tilde{\lambda}}}{(4\pi)^2}\Gamma\Big(-\frac{1}{2},\delta\Big)
+\frac{r_h\beta_0^2 {x}^2}{4\pi\sqrt{\tilde{\lambda}}}\Gamma\Big(\frac{1}{2},\delta\Big)\right],
\end{eqnarray}
where $ \delta= \frac{\tilde{\lambda} {x}^2}{4\pi\beta_0\epsilon}$ and $ \Gamma(z)=\int_{\delta}^{\infty} t^{z-1}e^{t}dt$ is incomplete Gamma function.

The $\epsilon$ is determined by the smallest length given by generalized uncertainty principle
\begin{equation}
\Delta X\Delta P=\frac{1}{2}e^{(\tilde{\lambda}(\Delta P)^2+\langle P\rangle)^2}.
\end{equation}
We can derive the least uncertainty of location 	$\sqrt{e\tilde{\lambda}/2}$. If we take it as a least length of pure space line element, we have
\begin{equation}
\label{elamb}
\sqrt{\frac{e\tilde{\lambda}}{2}}=\int_{r_{h}}^{r_h+\epsilon}\sqrt{g_{rr}}dr
\approx (1+\Theta)\int_{r_{h}}^{r_h+\epsilon}\frac{dr}{\sqrt{2\kappa(r-r_h)}}
=(1+\Theta)\sqrt{\frac{2\epsilon}{\kappa}}.
\end{equation}
Thus, from (\ref{elamb}), we have $ \delta= \frac{(1+\Theta)^2 {x}^2}{2\pi^2 e}$ and 
\begin{eqnarray}
S&=&\frac{(1-4\Theta)}{2\beta_0^2} \int^{\infty}_{0} 
\frac{ d{x}}{4\sinh^2({x}/2)}\left[\frac{\beta_0 {x}^4\sqrt{\tilde{\lambda}}}{(4\pi)^2}\Gamma\Big(-\frac{1}{2},\delta\Big)
+\frac{r_h\beta_0^2 {x}^2}{4\pi\sqrt{\tilde{\lambda}}}\Gamma\Big(\frac{1}{2},\delta\Big)\right],
\end{eqnarray}
with $ {x}\rightarrow 2{x} $, we have $ \delta= \frac{2(1+\Theta)^2 {x}^2}{\pi^2 e}$ and we obtain
\begin{eqnarray}
S&=&(1-4\Theta)\left[\frac{4\sqrt{\tilde{\lambda}}}{(4\pi)^2\beta_0}\delta_1
+\frac{r_h}{4\pi\sqrt{\tilde{\lambda}}}\delta_2\right]
\end{eqnarray}
being
 
\begin{eqnarray}
\delta_1=\int^{\infty}_{0} 
\frac{{x}^4}{\sinh^2({x})}\Gamma\Big(-\frac{1}{2},\delta\Big)d{x}, \quad\quad 
\delta_2=\int^{\infty}_{0}\frac{{x}^2}{\sinh^2({x})}\Gamma\Big(\frac{1}{2},\delta\Big)d{x},
\end{eqnarray}
when $ \sqrt{\tilde{\lambda}}=\delta_2/(2\pi^2) $, we find
\begin{eqnarray}
S&=\frac{1}{4}(1-4\Theta)(2\pi r_h)+\frac{(1-4\Theta)\delta_1\delta_2}{8\pi^4}T_h,
\end{eqnarray}
where $ 2\pi r_h $ is the horizon area of the noncommutatvie acoustic black hole. The second term is a correction term to the area entropy and is proportional to the radiation temperature of acoustic black hole.

\subsection{The case $B=0$ and $E\neq 0$}

In the present subsection we repeat the previous analysis for  $B=0$ and $E\neq 0$. As in the earlier case we take the acoustic line element obtained in \cite{ABP12}, in polar coordinates on the noncommutative plane, up to first order in $\theta$, in the { `non-relativistic'  limit} , given by
\begin{eqnarray}
\label{am}
ds^2&=&\left(1-\frac{3}{2}\theta\vec{\cal E}\cdot\vec{v}\right)
\left\{-[c^{2}-(v^2_{r}+v^2_{\phi}+\theta{\cal E}_{r}{v}_{r}+\theta{\cal E}_{\phi} {v}_{\phi})]dt^2
-2\left(v_{r}+\frac{\theta{\cal E}_{r}}{2}\right)drdt\right.
\nonumber\\
&-&\left.2\left(v_{\phi}+\frac{\theta{\cal E}_{\phi}}{2}\right)rd{\phi}dt
+(1-\theta{\cal E}_{r}{v}_{r}-\theta{\cal E}_{\phi} {v}_{\phi})(dr^2+r^2d\phi^2)\right\}.
\end{eqnarray}
where $ \theta\vec{\cal E}=\theta\vec{n}\times\vec{E} $, 
$ \theta{\cal E}_{r}= \theta(\vec{n}\times\vec{E})_{r}$ , 
$  \theta{\cal E}_{\phi}=\theta (\vec{n}\times\vec{E})_{\phi}$ and { $E$ is the magnitude of the electric field}.  Let us now consider the transformations of the time and the azimuthal angle coordinates as follows
\begin{eqnarray}
&&d\tau=dt+\frac{\tilde{v}_{r}dr}{(c^2-\tilde{v}_{r}^2)},
\nonumber\\
&&d\varphi=d\phi+\frac{\tilde{v}_{\phi}\tilde{v}_{r}dr}{r(c^2-\tilde{v}_{r}^2)}.
\end{eqnarray}
where we have defined $ \tilde{v}_{r}= v_{r}+\frac{\theta{\cal E}_{r}}{2}$ and $\tilde{v}_{\phi}=v_{\phi}+\frac{\theta{\cal E}_{\phi}}{2} $.
Now, we consider the flow with the velocity potential $\psi(r, \varphi) = A\ln{r}+ B\varphi$ whose velocity profile in polar coordinates on the plane is given by
$\vec{v}=\frac{A}{r}\hat{r}+\frac{B}{r}\hat{\phi}$.
Therefore, in these new coordinates the metric becomes
\begin{eqnarray}
\label{me}
ds^2\!&=&\!\left(1-\frac{3\theta{\cal E}_{r}A}{2r}-\frac{3\theta{\cal E}_{\phi} B}{2r}\right)\left\{
-\left[1-\frac{(A^2+B^2+\theta{\cal E}_{r}Ar+\theta{\cal E}_{\phi}Br)}{c^2r^2}\right]d\tau^2\right.
\nonumber\\
&+&\left.\left(1-\frac{\theta{\cal E}_{r}A}{r}-\frac{\theta{\cal E}_{\phi} B}{r}\right)\left[
\left(1-\frac{A^2+\theta{\cal E}_{r}Ar}{c^2r^2}\right)^{-1}dr^2+r^2d\varphi^2\right]
-2\left(\frac{B}{cr}+\frac{\theta{\cal E}_{\phi}}{2c}\right)rd\varphi d\tau\right\},
\end{eqnarray}
The radius of the ergosphere is given by $g_{00}(\tilde{r}_{e}) = 0$, whereas the horizon is given by the coordinate singularity $g_{rr}(\tilde{r}_{h}) = 0$, that is
\begin{eqnarray}\label{horizonR}
\tilde{r}_{e}=\frac{\theta{\cal E}_{r}A+\theta{\cal E}_{\phi} B}{2c^2}\pm\frac{1}{2}
\sqrt{\frac{(\theta{\cal E}_{r}A+\theta{\cal E}_{\phi} B)^2}{c^4}+4r^2_{e}}, 
\quad\quad \tilde{r}_{h_\pm}=
\frac{\theta{\cal E}_{r}A}{2c^2}\pm r_{h}\sqrt{1+\frac{(\theta{\cal E}_{r})^2}{4c^2}},
\end{eqnarray}
where $ r_{e}=\sqrt{(A^2+B^2)/c^2 }$  and $ r_{h}=|A|/c $ are the radii of the ergosphere and the horizon in the usual case. For $ \theta=0 $, we have  $\tilde{r}_{e}={r}_{e}$ and $\tilde{r}_{h}={r}_{h}  $.

Now we obtain the Hawking temperature of the acoustic
black hole as 
\begin{eqnarray}
T_{h}=\frac{k}{2\pi}=\left(1-\frac{3\theta{\cal E}_{r}}{2}\right)\frac{1}{2\pi r_{h}}+{\cal O}(\theta^2).
\end{eqnarray}
While the Unruh temperature for an observer at a distance $r$ is
\begin{eqnarray}
T=\frac{a}{4\pi}=\frac{f^{\prime}(r_h)}{4\pi}F^{-1/2}(r).
\end{eqnarray}
They satisfy the following relation
\begin{eqnarray}
T_h=\sqrt{F(r)}T=\frac{f^{\prime}(r_{h+})}{4\pi}.
\end{eqnarray}
where
\begin{eqnarray}
f(r)=\left(1-\frac{3\theta{\cal E}_{r}r_h}{2r}\right)\left(1-\frac{r^2_h}{r^2} -\frac{\theta{\cal E}_{r}r_h}{r}\right), 
\end{eqnarray}
and
\begin{eqnarray}
 F(r)&=&\frac{g_{tt}g_{\varphi\varphi}-g^2_{t\varphi}}{g_{\varphi\varphi}}
 \\
 &=&1-\frac{r^2_h}{r^2} -\theta\Big[\frac{1}{r}(4{\cal E}_r r_h +3{\cal E}_{\phi}B)
 -\frac{1}{r^3}\Big(3{\cal E}_r r_h^3 +3{\cal E}_{\phi}B r_h ^2+{\cal E}_rB^2 r_h +{\cal E}_{\phi}B^3 \Big)\Big].
\end{eqnarray}
Thus, near the horizon, $ \tilde{g}_{tt}(r_{h+})\rightarrow 0 $, we have $ {\cal E}_{\phi}B=-{\cal E}_{r}r_h $ and $ F(r) $ becomes 
\begin{eqnarray}
 F(r) &=&1-\frac{r^2_h}{r^2} -\frac{\theta}{r}{\cal E}_r r_h 
\end{eqnarray}
The entropy of the system, near the black hole horizon, is
\begin{eqnarray}
S&=&\frac{1}{2\beta_0^2} \int^{\infty}_{0} 
\frac{ dx}{4\sinh^2(x/2)}I(x,\epsilon),
\end{eqnarray}
where
\begin{eqnarray}
I(x,\epsilon)=\int\frac{\sqrt{g_{\varphi\varphi} g_{rr}}}{-\tilde{g}_{tt}}x^3e^{-\lambda\frac{x^2}{\beta^2}}dr
=\int\frac{x^3}{N^3}e^{-\lambda\frac{x^2}{\beta^2}}rdr, 
\end{eqnarray}
and 
\begin{eqnarray}
N^2=1-\frac{r^2_h}{r^2} -\frac{\theta}{r}{\cal E}_r r_h= 1-(1-\theta{\cal E}_r)\frac{r^2_{h+}}{r^2} -\frac{\theta}{r}{\cal E}_r r_{h+}
\end{eqnarray}
When $r \rightarrow r_{h+}$ , $N^2(r)\approx 2(1+\theta{\cal E}_r/2)\kappa (r-r_{h+}) $, so we have
\begin{eqnarray}
I(x,\epsilon)
=\int_{r_{h+}}^{r_{h+} +\epsilon}\frac{1}{(1+\frac{\theta{\cal E}_r}{2})^{3/2}}\frac{(r-r_{h+})+r_{h+}}{[2\kappa(r-r_{h+})]^{3/2}}x^3 
e^{-\tilde{\lambda} x^2/[2\kappa(r-r_{h+})\beta^2_0]}dr, 
\end{eqnarray}
{where $ \tilde{\lambda}=\lambda(1+\theta{\cal E}_r/2)^{-1} $ } and $ \kappa=2\pi\beta_0^{-1}$ is the surface gravity of acoustic black hole horizon and by variable substitution $ t=\frac{\tilde{\lambda} x^2}{4\pi(r-r_{h+})\beta_0} $, we have
\begin{eqnarray}
I(x,\epsilon)
&=&\frac{1}{(1+\theta{\cal E}_r/2)^{3/2}}\int_{\delta}^{\infty}\Big[\frac{\beta_0 x^4\sqrt{\tilde{\lambda}}}{(4\pi)^2}t^{-3/2} 
+\frac{r_{h+}\beta_0^2 x^2}{4\pi\sqrt{\lambda}}t^{-1/2}\Big]e^{-t}dt
\nonumber\\
&=&\frac{1}{(1+\theta{\cal E}_r/2)^{3/2}}\Big[\frac{\beta_0 x^4\sqrt{\tilde{\lambda}}}{(4\pi)^2}\Gamma\Big(-\frac{1}{2},\delta\Big)
+\frac{r_{h+}\beta_0^2 x^2}{4\pi\sqrt{\tilde{\lambda}}}\Gamma\Big(\frac{1}{2},\delta\Big)\Big],
\end{eqnarray}
where $ \delta= \frac{\tilde{\lambda} x^2}{4\pi\beta_0\epsilon}$ and $ \Gamma(z)=\int_{\delta}^{\infty} t^{z-1}e^{t}dt$ is incomplete Gamma function.

The $\epsilon$ is determined by the smallest length given by generalized uncertainty principle
\begin{equation}
\Delta X\Delta P=\frac{1}{2}e^{(\tilde{\lambda}(\Delta P)^2+\langle P\rangle)^2}.
\end{equation}
Again we can derive the least uncertainty of location $\sqrt{e\tilde{\lambda}/2}$. If we take it as a least length of pure space line element, we have
\begin{equation}
\label{elamb2}
\sqrt{\frac{e\tilde{\lambda}}{2}}=\int_{r_{h+}}^{r_{h+}+\epsilon}\sqrt{g_{rr}}dr
\approx \frac{1}{(1+\theta{\cal E}_r/2)^{1/2}}\int_{r_{h+}}^{r_{h+}+\epsilon}\frac{dr}{\sqrt{2\kappa(r-r_{h+})}}
=\frac{1}{(1+\theta{\cal E}_r/2)^{1/2}}\sqrt{\frac{2\epsilon}{\kappa}}.
\end{equation}
Thus, from (\ref{elamb}), we have $ \delta= \frac{(1+\theta{\cal E}_r/2)^{-1} x^2}{2\pi^2 e}$ and 
\begin{eqnarray}
S&=&\frac{1}{(1+\theta{\cal E}_r/2)^{3/2}}\frac{1}{2\beta_0^2} \int^{\infty}_{0} 
\frac{ dx}{4\sinh^2(x/2)}\Big[\frac{\beta_0 x^4\sqrt{\tilde{\lambda}}}{(4\pi)^2}\Gamma\Big(-\frac{1}{2},\delta\Big)
+\frac{r_{h+}\beta_0^2 x^2}{4\pi\sqrt{\tilde{\lambda}}}\Gamma\Big(\frac{1}{2},\delta\Big)\Big],
\end{eqnarray}
with $ x\rightarrow 2x $, we have $ \delta= \frac{2(1+\theta{\cal E}_r/2)^{-1} x^2}{\pi^2 e}$ and we obtain
\begin{eqnarray}
S&=&\frac{1}{(1+\theta{\cal E}_r/2)^{3/2}}\left[\frac{4\sqrt{\tilde{\lambda}}}{(4\pi)^2\beta_0}\delta_1
+\frac{r_{h+}}{4\pi\sqrt{\tilde{\lambda}}}\delta_2\right]
\end{eqnarray}
being
\begin{eqnarray}
\delta_1&=&\int^{\infty}_{0} 
\frac{x^4}{\sinh^2(x)}\Gamma\Big(-\frac{1}{2},\delta\Big)dx,
\quad\quad
\delta_2=\int^{\infty}_{0}\frac{x^2}{\sinh^2(x)}\Gamma\Big(\frac{1}{2},\delta\Big)dx,
\end{eqnarray}
when $ \sqrt{\tilde{\lambda}}=\delta_2/(2\pi^2) $, we find
\begin{eqnarray}
S&=\Big(1-\frac{3\theta{\cal E}_r}{4}\Big)\left[\frac{1}{4}(2\pi r_{h+})+\frac{\delta_1\delta_2}{8\pi^4}T_h\right],
\end{eqnarray}
where $ 2\pi r_{h+} $ is the horizon area of the noncommutative acoustic black hole. 
Note that, the correction term to the area entropy in 3-dimensional spacetime is proportional to the radiation temperature of acoustic black hole. For gravitational black hole in 4 dimensions the correction term is logarithmic. In our result in a 3-dimensional spacetime the logarithmic term does not exist. However, in this method, is not needed to introduce the ultraviolet cut-off and divergences are eliminated. Moreover, the small mass approximation is not necessary in the original brick-wall model.
Our calculations are consistent with the result obtained in~\cite{Zhao, Zhang}. 

{ 
It is also interesting to note from (\ref{elamb}) and (\ref{elamb2})  that the (increasing) decreasing of the least uncertainty of location as the noncommutative parameter increases is also responsible for (increasing) decreasing of the entropy. As a consequence, particularly in the former case, as long as the least uncertainty of location increases the entropy also increases. These parameters, of course, should be restricted in order to respect the second law of the entropy $\Delta S\geq0$. Other issues can also be addressed, e.g.,  the collapsing processes versus singularities, since collapsing processes tend to form singularities. In this sense the noncommutativity may play a role against the origin of singularities.
}

\acknowledgments

We would like to thank F.~G. Costa for fruitful discussions and CNPq, CAPES and PNPD/PROCAD -
CAPES for partial financial support.


\begin{thebibliography}{100}
\bibitem{Unruh} W. Unruh, Phys. Rev. Lett. {\bf 46}, 1351 (1981), -ibid, Phys. Rev. D {\bf 51}, 2827 (1995), [arXiv:gr-qc/9409008]; 

\bibitem{MV} M. Visser, Class. Quant. Grav. {\bf 15} 1767 (1998).

\bibitem{Volovik} G. Volovik, {\it The Universe in a Helium Droplet}, Oxford University Press, 2003;
L. C. B. Crispino, A. Higuchi, and G. E. A. Matsas, Rev. Mod. Phys. {\bf 80}, 787 (2008), [arXiv:0710.5373[gr-qc]];
 M. Cadoni, S. Mignemi, Phys. Rev. D {\bf 72}, 084012 (2005);
M. Cadoni, Class. Quant. Grav. {\bf 22}, 409 (2005).

\bibitem{others} L. C. Garcia de Andrade, Phys. Rev. D {\bf 70} (2004) 64004;
T. K. Das, {\it Transonic black hole accretion as analogue system}, [arXiv:gr-qc/0411006];
G. Chapline, P.O. Mazur, {\it Superfluid picture for rotating spacetimes},[arXiv:gr-qc/0407033];
O. K. Pashaev, J.-H. Lee, Theor. Math. Phys. 127 (2001) 779, [arXiv:hep-th/0104258];
S. E. Perez Bergliaffa, K. Hibberd, M. Stone, M. Visser, Physica D {\bf 191} (2004) 121,
[arXiv:cond-mat/0106255];
S. W. Kim, W.T. Kim, J.J. Oh, Phys. Lett. B {\bf 608} (2005) 10, [arXiv:gr-qc/0409003];
Xian-Hui Ge, Shao-Feng Wu, Yunping Wang, Guo-Hong Yang, You-Gen Shen,  Int. J. Mod. Phys. {\bf D21} (2012) 1250038, 
[arXiv:1010.4961 [gr-qc]];
C. Barcelo, S. Liberati, M. Visser, Living Rev. Rel. 8 (2005) 12, [arXiv:gr-qc/0505065];
E. Berti, V. Cardoso, J.P.S. Lemos, Phys. Rev. D {\bf 70} (2004) 124006, [arXiv:gr-qc/0408099];
V. Cardoso, J.P.S. Lemos, S. Yoshida, Phys. Rev. D {\bf 70} (2004) 124032, [arXiv:gr-qc/0410107];
Li-Chun Zhang, Huai-Fan Li, Ren Zhao, Phys. Lett. B {\bf 698} (2011) 438. 
C.~Furtado, A.~M.~M. de Carvalho, L.~C.~Garcia de Andrade and F.~Moraes,
  {\it Holonomy, Aharonov-Bohm effect and phonon scattering in superfluids,} gr-qc/0401025;
  F.~M.~Andrade, E.~O.~Silva and M.~Pereira,
  Phys.\ Rev.\ D {\bf 85}, 041701 (2012)  [arXiv:1112.0265 [quant-ph]];
 F. Correa, H. Falomir, V. Jakubsky, M. S. Plyushchay, J. Phys. A {\bf43} (2010) 075202,  [arXiv:0906.4055 [hep-th]]; 
  Francisco Correa, Horacio Falomir, Vit Jakubsky, Mikhail S. Plyushchay, 
Annals Phys. 325 (2010) 2653, [arXiv:1003.1434 [hep-th]];
Peter A. Horvathy, Mikhail S. Plyushchay.
Phys. Lett. B {\bf 595} (2004) 547, [hep-th/0404137];
Sergey M. Klishevich, Mikhail S. Plyushchay, Michel Rausch de Traubenberg.
Nucl. Phys. B 616 (2001) 419, [hep-th/0101190].

\bibitem{RS} R. Sch\"{u}tzhold and W. G. Unruh, Phys. Rev. D {\bf 66}, 044019 (2002).
\bibitem{Mathis} G. Rousseaux, C. Mathis, P. Ma\"{i}ssa, T. G. Philbin, and U. Leonhardt, New Journal of Physics {\bf 10}, 053015 (2008).
\bibitem{UL} U. Leonhardt and P. Piwnicki, Phys. Rev. Lett. {\bf 84}, 822 (2000); U. Leonhardt, Nature {\bf 415}, 406 (2002); 
W. G. Unruh and R. Sch\"{u}tzhold, Phys. Rev. D {\bf 68}, 024008 (2003).
\bibitem{Philbin} T. G. Philbin, C. Kuklewicz, S. Robertson, S. Hill, F. K\"{o}nig, and U. Leonhardt, Science {\bf 319}, 1367 (2008).
\bibitem{RSch} R. Sch\"{u}tzhold and W. G. Unruh, Phys. Rev. Lett. {\bf 95}, 031301 (2005).
\bibitem{Novello} M. Novello, M. Visser and G. Volovik (Eds.), {\it Artificial Black Holes}, World Scientific, Singapore, (2002).
\bibitem{Garay} L. J. Garay, J. R. Anglin, J. I. Cirac, and P. Zoller, Phys. Rev. Lett. 85, 4643 (2000).

\bibitem{OL} O. Lahav, A. Itah, A. Blumkin, C. Gordon and J. Steinhauer, [arXiv0906.1337].
\bibitem{SG} S. Giovanazzi, Phys. Rev. Lett. {\bf 94}, 061302 (2005).

\bibitem{Xian} X.-H. Ge, S.-J. Sin, JHEP {\bf 1006}, 087 (2010) [arXiv:hep-th/1001.0371]; 
N. Bilic, Class. Quant. Grav. {\bf16} (1999) 3953; 
S. Fagnocchi, S. Finazzi, S. Liberati, M. Kormos and A. Trombettoni [arXiv:1001.1044[gr-qc]];
M. Visser and C. Molina-Paris, [arXiv:1001.1310[hep-th]].


\bibitem{ABP} M. A. Anacleto, F. A. Brito, E. Passos, Phys. Lett. B {\bf694}, 149 (2010)  [arXiv:1004.5360 [hep-th]]; Phys. Lett. B {\bf703}, 609 (2011) [arXiv:1101.2891 [hep-th]];
Phys. Rev. D {\bf 86}, 125015 (2012)  [arXiv:1208.2615 [hep-th]]; {\it Acoustic Black Holes and Universal Aspects of Area Products }  [arXiv:1309.1486] .

\bibitem{Rinaldi}  S. Giovanazzi, Phys. Rev. Lett. 106, 011302 (2011), M. Rinaldi, Phys. Rev. D {\bf 84}, 124009 (2011) .

\bibitem{Rinaldi:2011aa}
   M.~Rinaldi,
   Int.\ J.\ Mod.\ Phys.\ D {\bf 22} (2013) 1350016
   [arXiv:1112.3596 [gr-qc]].

\bibitem{Wilczek} V. Frolov, I. Novikov, Phys. Rev. D {\bf 48}, 4545 (1993) ; 
C.G. Callan, F. Wilczek, Phys. Lett. B {\bf 33}, 55 (1994).

\bibitem{XLi} X. Li, Z. Zhao, Phys. Rev. D {\bf 62}, 104001 (2000);
R. Zhao, S. L. Zhang, Gen. Relat. Grav. {\bf 36}, 2123 (2004);
R. Zhao, Y. Q. Wu, L. C. Zhang, Class. Quantum Grav. {\bf 20}, 4885 (2003);
W. Kim, Y. W. Kim, Y. J. Park, Phys. Rev. D {\bf 74}, 104001 (2006); Phys. Rev. D {\bf 75}, 127501 (2007);
M. Yoon, J. Ha, W. Kim, Phys. Rev. D {\bf 76},  047501 (2007);
Y. W. Kim, Y. J. Park, Phys. Lett. B {\bf 655}, 172 (2007).

\bibitem{KN} K. Nouicer, Phys. Lett. B {\bf 646}, 63 (2007).

\bibitem{Zhao} HuiHua Zhao, GuangLiang Li, LiChun Zhang, Phys.Lett. A {\bf 376},  2348 (2012).

\bibitem {Zhang} R. Zhao, L.C. Zhang, H.F. Li, Acta Phys. Sin. 58,  2193 (2009) . 

\bibitem{ABP12} M. A. Anacleto, F. A. Brito, E. Passos, Phys. Rev. D {\bf 87}, 125015 (2013) [ arXiv:1210.7739 ]; Phys. Rev. D {\bf 85}, 025013 (2012) [arXiv:1109.6298 [hep-th]], 

\bibitem{shk} 
  J.~Casalderrey-Solana, E.~V.~Shuryak and D.~Teaney,
  J.\ Phys.\ Conf.\ Ser.\  {\bf 27}, 22 (2005)
  [Nucl.\ Phys.\  A {\bf 774}, 577 (2006)]
  [arXiv:hep-ph/0411315].
\bibitem{BH-plasma} 
  L.~C.~Garcia de Andrade,
  ``Kerr-Schild Riemannian acoustic black holes in dynamo plasma laboratory,''
  arXiv:0808.2271 [gr-qc].

\bibitem{syb}
  S.~L.~Dubovsky and S.~M.~Sibiryakov,
  Phys.\ Lett.\  B {\bf 638}, 509 (2006)
  [arXiv:hep-th/0603158].


\bibitem{adam}
  E.~Kant, F.~R.~Klinkhamer and M.~Schreck,
  Phys.\ Lett.\  B {\bf 682}, 316 (2009)
  [arXiv:hep-th/0909.0160].

\bibitem{Basak:2002aw}
  S.~Basak and P.~Majumdar,
  Class.\ Quant.\ Grav.\  {\bf 20}, 3907 (2003)
  [arXiv:gr-qc/0203059].

\bibitem{SBP}S. Basak and P. Majumdar, Class. Quant. Grav. {\bf 20}, 2929 (2003); 
S. Basak, \textit{Analog of superradiance effect in BEC} [arXiv:gr-qc/0501097]; S. Basak, \textit{Sound wave in vortex with sink} [arXiv:gr-qc/0310105];
M. Richartz, S. Weinfurtner, A. J. Penner, W. G. Unruh, Phys. Rev. D {\bf 80}, 124016 (2009), [arXiv:0909.2317 [gr-qc]]; 
S. Lepe, J. Saavedra, Phys. Lett. B617, 174 (2005).

\bibitem{SW} N. Seiberg and E. Witten, JHEP {\bf 09} 032 (1999),  [hep-th/9908142].

\bibitem{SGhosh} S. Ghosh, Mod. Phys. Lett. A{\bf20}, 1227 (2005).

\bibitem{rivelles} V.O. Rivelles, Phys. Lett. B {\bf 558} 191 (2003),  hep-th/0212262; V.O. Rivelles, PoS WC2004, 029 (2004),  
hep-th/0409161;
T. Mariz, J.R. Nascimento, V.O. Rivelles, Phys. Rev. D {\bf} 75, 025020 (2007),  hep-th/0609132.

\bibitem{revnc} For a review on noncommutative field theories see: R. J. Szabo, Phys. Rept. {\bf 378}, 207 (2003),  
[arXiv:hep-th/0109162]; 
M. R. Douglas and N. A. Nekrasov, Rev. Mod. Phys. {\bf 73}, 977 (2001),  [arXiv:hep-th/0106048]; 
V. O. Rivelles, Braz. J. Phys. {\bf 31}, 255 (2001)  [arXiv:hep-th/0103131]; 
M. Gomes, {\it in Proceedings of the XI Jorge Andr\'e Swieca Summer School, Particles and Fields}, edited by
G. A. Alves, O. J. P. \'Eboli, and V. O. Rivelles (World Scientific, Singapore, 2002); H. O. Girotti,  hep-th/0301237; Piero Nicolini, Int. J. Mod. Phys. A24, 1229 (2009) , [arXiv:0807.1939 [hep-th]].
  R.~Casana and K.~A.~T.~da Silva,
  [arXiv:1106.5534 [hep-th]];
A. Kobakhidze and B. H. J. McKellar,
   Phys. Rev. D {\bf 76}, 093004 (2007),
   [arXiv:0707.0343 [hep-ph]].
  R.~Banerjee, B.~Chakraborty, S.~Ghosh, P.~Mukherjee and S.~Samanta,
  Found.\ Phys.\  {\bf 39}, 1297 (2009)
  [arXiv:0909.1000 [hep-th]].
  
\bibitem{Cai:2007xr}  
  Y.~-f.~Cai and Y.~-S.~Piao,
  Phys.\ Lett.\ B {\bf 657}, 1 (2007)
  [gr-qc/0701114].
  
\bibitem{Cai:2014hja} 
  Y.~-F.~Cai and Y.~Wang,
  arXiv:1404.6672 [astro-ph.CO].

\end{thebibliography}
\end{document}